\documentclass[12pt,a4paper]{article}
\newcommand{\W}{{\cal W}}
\newcommand{\Tr}{{\rm Tr}}
\newcommand{\gSquare}{g^2}
\newcommand{\gSOne}{g^2}
\newcommand{\gCube}{g^3}
\newcommand{\gCOne}{g^3}

\begin{document}

\begin{center}
{\bf QCD partition function in the external field in the covariant gauge}
\end{center}

\bigskip

\begin{center}
{\bf Andrei Leonidov, Vladimir Nechitailo}
\end{center}

\begin{center}

\medskip

{\it  Theoretical Physics Department, P.N. Lebedev Physics Institute,

      119991 Leninsky pr. 53, Moscow, Russia}

and

{\it  Institute of Theoretical and Experimental Physics

117259 B. Cheremushkinskaya 25, Moscow, Russia}
\end{center}

\bigskip

\begin{abstract}
The QCD partition function in the external stationary gluomagnetic field is computed in the third order in external field invariants in arbitrary
dimension and arbitrary covariant gauge. The contributions proportional to third order invariants in gluon field strength are shown to be dependent
on covariant quantum gauge fixing parameter $\alpha$.
\end{abstract}

\newpage

\section{Introduction}

The notion of effective action plays a central role in exploring the nonlinear effects in quantum field theory \cite{S89}. This language especially
convenient for analyzing the situations in which a description can be separated into contributions of quantum modes and external fields. Taking into
account quantum fluctuations changes the tree-level description of the system through new nonlinear contributions in the tree-level fields that
effectively correspond to new terms in the action of the system. Technically the computation is done for quantum modes propagating in the
corresponding external field. These computations are, in particular in Yang-Mills theory, rather nontrivial. The physical answer should ideally be
explicitly gauge invariant with respect to gauge transformations of both tree-level and quantum gauge field configurations. The invariance with
respect to gauge transformations of classical fields is ensured by using the background field gauge \cite{DW81}. Much more difficult is to cope with
the dependence on the quantum gauge fixing parameter. Generically nothing forbids a dependence of contributions proportional to higher order
invariants in the external field being from the quantum gauge fixing procedure chosen. Explicit calculations of corresponding contributions in high
energy QCD \cite{MQ86,AL00} illustrate this point. An interesting suggestion of coping with such problems is to consider a so-called gauge invariant
effective action \cite{VDW84}, which in the case of Yang-Mills theory amounts to performing calculations in the Landau gauge $\alpha=0$.

Of special interest is calculation of the effective action on manifolds with cylindrical geometry corresponding to computing the partition function
(free energy) of the theory under consideration at finite temperature. The calculation of the one-loop QCD partition function in external
gluomagnetic field including the third order invariants in the external gluon field strength in the Feynman gauge $\alpha=1$ was performed in
\cite{L90a}. Fully resummed nonlocal contributions to a number of lowest order contributions was computed in \cite{LZ92}. Let us note that apart from
having purely theoretical significance, analysis of nonlinear contributions to the effective action (free energy) at finite temperature is of clear
phenomenological interest. Nonlinear terms lead, in particular, to modification of the speed of sound in the nonabelian medium \cite{L90b} which
affects its hydrodynamical evolution \cite{N94}.

In the present paper we generalize the calculation of \cite{L90a} to
the case of arbitrary dimension and arbitrary covariant gauge.

\section{Computation of the partition function}

The expression for the one-loop partition function (Z) of gluodynamics in
the covariant gauge reads
\begin{eqnarray}\label{pf}
    - \frac{F}{T}  =\ln \frac{Z[A,T]}{Z[0,T]} & = & -\frac{1}{4T}\int d^{2\omega}x (G^{a}_{\mu\nu})^2 - \frac{1}{2}\ln \det \W(\alpha) + \ln \det ( - \frac{1}{\sqrt{\alpha}} D^2) \nonumber\\
    &\equiv& -\frac{1}{4T}\int d^{2\omega}x (G^{a}_{\mu\nu})^2 - \left( \frac{F}{T} \right)^{gl} - \left( \frac{F}{T} \right)^{gh}
\end{eqnarray}
Here $F$ is free energy, $T$ is the temperature, $D^2$ is a kinetic operator for the ghosts, and $\W^{ab}_{\mu\nu}$ is a gluon kinetic operator
\begin{equation}
\W^{ab}_{\mu\nu}= -D^{2}(A)^{ab}\delta_{\mu\nu}-2f^{abc}G^{c}_{\mu\nu}- (\frac{1}{\alpha}-1)D^{ac}_{\mu}D^{cb}_{\nu}
\end{equation}
where $\alpha$ is a gauge fixing parameter, $f^{abc}$ are the structure constants of the corresponding Lie algebra, $D^a_{\mu}$ is a covariant
derivative containing the external gauge field potential $A^a_{\mu}$ and $G^a_{\mu \nu}$ is the corresponding field strength. Let us emphasize that
we keep in (\ref{pf}) a manifest dependence on the gauge parameter $\alpha$-dependence in the ghost determinant (usually it is hidden in the
normalization constant) because of its importance in performing calculations in the Landau gauge $\alpha\rightarrow 0$.

In computing the
logarithms of the determinants in (\ref{pf}) it is convenient to use a
$\zeta$ - regularization
\begin{equation}\label{zeta}
\ln \det A = - \lim_{s\rightarrow 0} \frac{d\ }{ds}\frac{M^{2s}}{\Gamma(s)} \int^{\infty}_{0} dt\,t^{s-1}\Tr e^{-t A}
\end{equation}
As the closed expression for the right-hand side of (\ref{zeta}) does
not exist, one is forced to use an expansion of $\Tr e^{-t A}$ in the
invariants of the background field (Seeley expansion) \cite{S68}. For
example, at zero temperature the expansion of gluon kinetic operator ${\cal W}$ reads
\cite{RS79}:
\begin{equation}\label{seeexp1}
  {\rm Tr} \langle x |  e^{-\W^{ab}_{\mu \nu} s} | x \rangle  \sim
  (\int d^{2\omega}x) \sum_{k=0}^{\infty} b_{-w+k} (G_{\mu \nu},w,\alpha |  x )  s^{-w+k}
\end{equation}
To calculate the functional trace in (\ref{seeexp1})  it is convenient to use the basis of plane waves (see, e.g., \cite{GolA52},\cite{DPYu84}):
\begin{eqnarray}\label{TrW}
\Tr\ <x|e^{-\W s}|x> &=& \Tr  \int \frac{d^{2\omega}p}
 {(2\pi)^{2\omega}}\ e^{-ipx}
<x|e^{-{\cal W}s}|x>e^{ipx} \nonumber\\
& = &\Tr \int d^{2\omega}x \int \frac{d^{2\omega}p}{(2\pi)^{2\omega}}
e^{-s{\cal W}( \partial_{\mu} \rightarrow \partial_{\mu} + ip_{\mu} )}1 \, ,
\end{eqnarray}
where the last trace is performed over the Lorenz and color
indices.
The explicit expressions for the first four Seeley coefficients at zero temperature were
computed in \cite{GLNO94,GLNO96} ( we refer for the detailed description of the calculation
method to these papers)
\begin{eqnarray}\label{sc0123}
b_{-\omega\phantom{+0}} &=&\frac{N^2_c-1}{2^{2\omega}\pi^{\omega}}
              (2\omega - [1-\alpha^{\omega}]),
\nonumber \\
b_{-\omega+1} &=& \frac{N_c}{2^{2\omega}\pi^{\omega}}
              (2\omega - [1-\alpha^{\omega-1}])
 \frac{\Gamma(\omega)-\frac{1}{\omega}\Gamma(\omega+1)}{\Gamma(\omega)}
 A^a_\mu A^a_\mu \equiv 0 ,
\nonumber  \\
b_{-\omega+2} &=& \frac{N_c}{2^{2\omega}\pi^{\omega}}
              \{ 2 -\frac{1}{12} (2\omega - [1-\alpha^{\omega-2}]) \} G_2
                .
\\
b_{-\omega+3} &=& - \frac{N_c}{2^{2\omega}\pi^{\omega}}
 \left[\frac{1}{180}(2\omega - [1-\alpha^{\omega-3}])(G_3 - 3 I_3)
 + \frac{2}{3}I_3 + \xi(\omega,\alpha) I_3 \right] \nonumber
\end{eqnarray}
where
\begin{equation}\label{Ginv}
G_2 \equiv G^a_{\mu\nu} G^a_{\mu\nu}, \quad
G_3 \equiv f^{abc} G^a_{\mu \nu} G^b_{\nu \rho} G^c_{\rho \mu},  \quad
I_3 \equiv (D^{ab}_{\mu} G^b_{\mu \nu}) (D^{ac}_{\rho} G^c_{\rho \nu})
\end{equation}
are the invariants one can construct from gluon field strength
tensor and
\begin{eqnarray*}\label{ksif}
\xi(\omega,\alpha)\equiv
 \frac{1}{4(\omega-1)} \left\{
 \frac{1}{\omega} \left( 4\omega -1-\alpha
 - \frac{\alpha^2}{1-\alpha} [1-\alpha^{\omega-2}] \right)
\right.
\\
\left.
 - \frac{3\alpha}{1-\alpha}\frac{1-\alpha^{\omega-2}}{\omega-2}
 \right\} - \frac{1-\alpha^{\omega-2}}{6(\omega-2)}
\end{eqnarray*}
represents nontrivial off-shell contribution vanishing in the Feynman gauge at $\alpha=1$.

To calculate the trace in (\ref{TrW}) at the finite temperature it is convenient to use the Matzubara formalism in which the Eucledian space has
cylindrical geometry. The extension of the compact dimension is the inverse temperature $\beta=1/T$ and the fields are periodic in this coordinate
with the period $\beta$. As a result the temporal integral in (\ref{seeexp1}) is replaced by the sum over Matsubara frequencies $\omega_{n} =
(2\pi/\beta) n$ and (\ref{TrW}) takes the following form
\begin{eqnarray}\label{TrWbeta}
\Tr <x|e^{-\W s}|x> &=&
Tr \int\limits^{\beta}_{0}d \tau \int d^{2\omega-1}x \frac{1}{\beta}
 \sum^{\infty}_{n=-\infty} \int \frac{d^{2\omega-1}p}{(2\pi)^{2\omega-1}}
\nonumber \\
&&
\times e^{-s\W(\partial_{0} \rightarrow \partial_{0} + i\omega_{n}, \partial_{j} \rightarrow \partial_{j} + ip_{j}
 )}1
\end{eqnarray}
Since the case of zero Matsubara frequency $n=0$ is completely analogous to
the zero temperature case for the theory in dimension $2\omega-1$ in what follows we consider only the temperature-dependent contributions
$ \sum^{\infty}_{n=-\infty}  \rightarrow 2\sum_{n = 0}^{\infty} $ and keep only even powers of $\omega_{n}$.

Technically summation in (\ref{TrWbeta}) can be performed  in closed form only together with integration
over proper time in (\ref{zeta}).

Let us first consider the contributions to the free energy in the
first nontrivial order $D^4$. For the gluonic contribution one has\footnote{In what follows we assume that invariants (\ref{ksif}) include integration over corresponding spatial coordinates}
\begin{eqnarray}\label{Fgl4}
-\left( \frac{F}{T} \right)^{gl}_4 & = & \frac{1}{2}\lim\limits_{s\rightarrow 0}\frac{d}{ds}
\frac{\Gamma(s+1/2+2-\omega)}{8\pi^{4-\omega}\sqrt{\pi}}
\frac{\zeta(2s+5-2\omega)}{\Gamma(s)} \\ & \times & \left(\frac{M^2}{4\pi^2 T^2}\right)^s
\left[ 2-\frac{2\omega-1+\alpha^s}{12}\right] \frac{\gSquare N_c}{T^{5-2\omega}}  G_2 \nonumber
\end{eqnarray}
The ghost contribution in the same order $D^4$ reads
\begin{eqnarray}\label{Fgh4}
 -\left( \frac{F}{T} \right)^{gh}_4 & = &  \lim\limits_{s\rightarrow 0}\frac{d}{ds}
\frac{\Gamma(s+1/2+2-\omega)}{8\pi^{4-\omega}\sqrt{\pi}}
\frac{\zeta(2s+5-2\omega)}{\Gamma(s)} \\ & \times & \left(\frac{M^2\sqrt{\alpha}}{4\pi^2 T^2}\right)^s
 \frac{1}{12} \frac{\gSquare N_c}{T^{5-2\omega}}  G_2 \nonumber
\end{eqnarray}
The contributions to the free energy of gluons and ghosts in the next
order $D^6$ read (here when taking a limit $s\rightarrow 0$ we assume that $\omega \neq 3$)
\begin{eqnarray}\label{Fgl6}
 -\left( \frac{F}{T} \right)^{gl}_6 & = & - \frac{\zeta(7-2\omega)\Gamma(1/2+3-\omega)}{2^6\pi^{6-\omega}\sqrt{\pi}} \\
& \times &  \frac{\gCube N_c}{T^{7-2\omega}}
 \left\{\frac{2\omega}{180}(G_3-3I_3) + \frac{2}{3}I_3 + (1-\alpha)\frac{7+\alpha}{24}I_3\right\} \nonumber
\end{eqnarray}
and
\begin{equation}\label{Fgh6}
 -\left( \frac{F}{T} \right)^{gh}_6  =
 \frac{\zeta(7-2\omega)\Gamma(1/2+3-\omega)}{2^6\pi^{6-\omega}\sqrt{\pi}}
\frac{\gCube N_c}{T^{7-2\omega}} \left\{\frac{2}{180}(G_3-3I_3)\right\}
\end{equation}
Combining the gluon and ghost contributions for the free energy
Eqs.~(\ref{Fgl4})-(\ref{Fgh6}) we obtain the final expression for the
regularized free energy
\begin{eqnarray}\label{Fun}
- \left( \frac{F}{T} \right)^{gl} - \left( \frac{F}{T} \right)^{gh} & = & \lim\limits_{s\rightarrow 0}\frac{d}{ds} \frac{\Gamma(s+1/2+2-\omega)}{16\pi^{4-\omega}\sqrt{\pi}}
\frac{\zeta(2s+5-2\omega)}{\Gamma(s)} \\
& \times & \left(\frac{M^2}{4\pi^2 T^2}\right)^s \left[ 2-\frac{2\omega-1+\alpha^s - 2\alpha^{s/2}}{12}\right] \frac{\gSquare N_c}{T^{5-2\omega}}  G_2 \nonumber \\
& - & \frac{\zeta(7-2\omega)\Gamma(1/2+3-\omega)}{2^6\pi^{6-\omega}\sqrt{\pi}}
\frac{\gCube N_c}{T^{7-2\omega}} \nonumber \\
& \times &
 \left\{\frac{2 (\omega-1)}{180}(G_3-3I_3) + \frac{2}{3}I_3 +
 (1-\alpha)\frac{7+\alpha}{24}I_3\right\} \nonumber
\end{eqnarray}

We see that in the last formula the $\alpha$ and $\omega$ dependences are factorized
unlike the zero temperature case (\ref{sc0123}).

Special care has to be exercised when computing the QCD partition
function in the physical dimension $d=4$ ($\omega=2$). Let us first
consider the contributions of the fourth order in covariant
derivatives. The expressions for the gluon and ghost contribution to
the free energy read

\begin{equation}\label{Fgl4d2}
-\left( \frac{F}{T} \right)^{gl}_4 \, = \,  \frac{\gSOne}{16\pi^2}\frac{N_c}{3}
\left(10 \ln\frac{M^2}{T^2}-\frac{1}{2}\ln\alpha\right) \frac{1}{4T}  G_2
\end{equation}
and

\begin{equation}\label{Fgh4d2}
-\left( \frac{F}{T} \right)^{gh}_4 \, = \, \frac{\gSOne}{16\pi^2}\frac{N_c}{3}
\left(\ln\frac{M^2}{T^2}+\ln\sqrt{\alpha}\right) \frac{1}{4T} G_2 \, .
\end{equation}
%
We see that the expressions Eqs.~(\ref{Fgl4d2}) and (\ref{Fgh4d2}) contain contributions proportional to the logarithm of the quantum gauge fixing
parameter $\log \alpha$ which are singular in the limit $\alpha \to 0$. However, these terms in Eqs.~(\ref{Fgl4d2}) and (\ref{Fgh4d2}) cancel each
other upon adding the gluon and ghost contributions. There are no other contributions singular in $\alpha$ in the next order in covariant derivatives
in any dimentions\footnote{At $\omega=3$ we have no singularities in $\alpha$  in the order $D^4$ and the same logarithmic cancellation in the order
$D^6$ as for $\omega=2$ in the order $D^4$}. Let us emphasize again that all singularities in $\alpha$ disappear only after regularization and taking
into account the ghost contribution.

At zero temperature we have the leading singularites of the form $[1-\alpha^{\omega-N}]$ and "subleading" terms which come from the contribution of
order  $D^6$ in Eq.~(\ref{ksif}) that in the limit of $\omega \to 2$ take the following form:

\begin{equation}\label{ksifd2}
\xi(2,\alpha) \, =  \, \frac{1}{24}\{ 3(7-\alpha) + 2 \frac{\ln(\alpha)}{(1-\alpha)}(2+7\alpha) \}.
\end{equation}
This basic difference between the zero and finite temperature cases is due to the existence of the natural energy scale ($T$) in the last case. If we
try to estimate the corresponding determinants following \cite{DPYu84} by introducing
 the energy scale $\Lambda^{-1} = - b_0/b_1$ and writing
\begin{equation}
\ln \det \W = - b_0( \ln \frac{M^2}{\Lambda} + \gamma_E - 1 )
\label{ApprDet}
\end{equation}
($\gamma_E$ is Euler's constant) we will have the same cancellation of $\alpha \to 0$ singularity. To elucidate this statement let us consider a
variation of the functional determinant under scale transformation. By using $\zeta$-function definition of a determinant it is easy to get the
following equation \cite{S89}:

\begin{equation}
 \ln \det(\frac{1}{\alpha}B) =
\ln \det(B) - b_0\ln \alpha
\label{det}.
\end{equation}

If now we estimate both of the above determinants by using the formula (\ref{zeta}) directly we obtain, instead of (\ref{det}), the following
expression:

\begin{displaymath}
\sum^{\infty}_{ N=1}
(1-\alpha^{\omega-N}) \Lambda^{\omega-N} b_{-\omega+N} \approx
 - b_0\ln \alpha
\end{displaymath}

\noindent where $1/\Lambda$ is the (upper) cutoff parameter in the integral over proper time. By comparing this expression with the explicit form of
the Seeley coefficients (\ref{sc0123}) we conclude that all $[1-\alpha^{\omega-N}]$-terms give nothing else than the contribution $ - b_0 \ln \alpha
$ in the final answer.

Summing all
contributions to the free energy we obtain the final expression for
the logarithm of the QCD partition function in the physical dimension
$d=4$ ($\omega=2$):
\begin{eqnarray}\label{F462d}
\ln \frac{Z[A,T]}{Z[0,T]} &=& -\left( 1-\frac{\gSOne}{16\pi^2}\frac{N_c}{3}
11 \ln\frac{M^2}{T^2}\right)  \frac{1}{4T}  G_2
 \\
&&-\zeta(3)\frac{\gCOne}{64\pi^4}\frac{N_c}{T^3}\left\{  \frac{1}{180}(G_3-3I_3)+\left(\frac{1}{3}+\frac{1-\alpha}{48}(7+\alpha)\right)I_3\right\}\nonumber
\end{eqnarray}

Let us stress that although the formula (\ref{F462d}) does not contain singularities in the quantum gauge fixing parameter $\alpha$, the answer does depend on it off-shell.

Expressions in Eqs.~(\ref{Fun}) and (\ref{F462d}) constitute the main results of the paper.

\section{Conclusion and outlook}

Let us formulate the main result of the present paper. Using the heat kernel expansion we have computed the partition function of pure Yang-Mills theory to the third order in the invariants in the external field strength in arbitrary dimension and arbitrary covariant gauge in the case when only stationary spatial ("magnetic") gauge potentials of the external field were taken into account. The calculation demonstrates delicate $\alpha$ - dependent cancellations between the gluon and ghost contributions to the partition function. The first nontrivial nonlinear contribution to the effective action was shown to depend on the quantum gauge fixing parameter $\alpha$.

The major question left unanswered at this stage is to complete the program by calculating the partition function taking into account the temporal gauge connection $A_0$. Work in this direction is currently in progress \cite{LN05}.

\bigskip

{\bf Acknowledgments.} The work of A.L. was partially supported by the RFBR Grant 04-02-16880, and the Scientific school support grant 1936.2003.02,
the work of V.N. was supported in part by the RFBR grant 04-02-16445, Russian Science Support Foundation, and Dynasty Foundation.


\begin{thebibliography}{99}


\bibitem{S89}
A.S.~Shwarz, {\it Quantum Field Theory and Topology}, Moskva, Nauka, 1989 (in Russian).


\bibitem{DW81}
B.S. DeWitt, "A Gauge Invariant Effective Action",
              in {\it Quantum Gravity II}, eds. C.J. Isham,
              R. Penrose and D.W. Sciama, (~Oxford Univ. Press,
              Oxford 1981~).

\bibitem{MQ86}
A.H.~Mueller, J.W.~Qui, {\it Nucl.\ Phys.}\ {\bf 268} (1986), 427

\bibitem{AL00}
A.~Leonidov, in {\it Quantization, Gauge Theory and Strings: Proc. of the Intern. Conf. Dedicated to the Memory of Professor Efim Fradkin, Moscow,
Russia, June 5-10, 2000.} Vol. 2. (Eds. A.~Semikhatov, M.~Vasiliev, V.~Zaikin.) (Moscow: Scientific World, 2001) p.542

\bibitem{VDW84}
G.A. Vilkovisky, {\it Nucl.\, Phys.}\, {\bf B234} (1984), 125;\\
              B.S. DeWitt, "The Effective Action",
              in {\it Quantum Field Theory and Quantum Statistics:
              Essays in Honour of the Sixtieth Birthday of E.S. Fradkin},
              Eds. I.A. Batalin, C.J. Isham and G.A. Vilkovisky
              (Adam Hilger: Bristol 1987).

\bibitem{L90a}
A.V.~Leonidov, {\it Z.\ Phys.}\ {\bf C47} (1990), 287-289

\bibitem{LZ92}
A.~Leonidov, A.~Zelnikov, {\it Phys.\ Lett.}\ {\bf 276} (1992)

\bibitem{L90b}
A.V.~Leonidov, "On Power Corrections in Finite Temperature QCD", Vacuum structure in intense fields, Cargese 1990, Proceedings,  pp. 373-376.

\bibitem{N94}
V.A.~Nechitailo,
{\it Phys. Atom. Nucl.} {\bf 57} (1994) 1315.

\bibitem{S68}
R.T.~Seeley, {\it Mathematika}\ {\bf 12} (1968)

\bibitem{RS79}
V.N.~Romanov, A.S.~Schwartz, {\it Teor. Mat. Fiz.}\, {\bf 41} (1979), 190

\bibitem{GolA52}
M.L.Goldberger, E.N.Adams, {\it Journ.Chem.Phys.}\ {\bf 20} (1952), 240.

\bibitem{DPYu84}
D.I.~D'yakonov, V.Yu.~Petrov and A.V.~Yung. {\it Sov.J.Nucl.Phys.} {\bf 39}(1984) 150.

\bibitem{GLNO94}
E.I.~Guendelman, A.V.~Leonidov, V.A.~Nechitailo and D.A.~Owen, {\it Phys.\ Lett.}\, {\bf B234} (1994), 160

\bibitem{GLNO96}
E.I.~Guendelman, A.V.~Leonidov, V.A.~Nechitailo and D.A.~Owen, {\it J.\ Phys.}\, {\bf A29} (1996), L211-215

\bibitem{LN05}
A.~Leonidov, V.~Nechitalo, work in progress.


\end{thebibliography}
\end{document}